\documentclass{aa}  
\usepackage{graphicx}
\usepackage{txfonts}
\usepackage{natbib}
\bibpunct{(}{)}{;}{a}{}{,}
\begin{document}
   \title{Anomalous X-Ray emission in GRB\,060904B: a Nickel line? }


   \author{R. Margutti
          \inst{1,2}
          \and
          A. Moretti\inst{2}
	  \and
	  F. Pasotti
	  \inst{2}
	  \and
	  S. Campana
	  \inst{2}
	  \and
	  G. Chincarini 
	\inst{1,2}
	\and
	S. Covino
	\inst{2}
	\and
	C. Guidorzi
	\inst{1,2}
	\and
	P. Romano
	\inst{1,2}
	\and
	G. Tagliaferri
	\inst{2}
   }

   \offprints{R. Margutti}

   \institute{Universit\'a degli Studi di Milano-Bicocca, Dipartimento di Fisica, Piazza della Scienza 3, I-20126 Milano, Italy\\
              \email{raffaella.margutti@brera.inaf.it}
         \and INAF, Osservatorio Astronomico di Brera, Via E. Bianchi 46, I-23807, Merate (LC), Italy\\
             \email{}
             }

   \date{Received ??; accepted ??}

 
  \abstract
   {The detection of an extra component in GRB\,060904B X-ray spectra in
	addition to the standard single power-law behaviour has recently been reported in the literature.
	This component can be fit with different models; in particular the addition of a spectral line
	provides the best representation.} 
   {In this paper we investigate the physical properties that the surrounding medium
must have in order to produce a spectral feature that can explain the detected emission.} 
   {We analyse and discuss how and if the detected spectral excess fits in different theoretical
models developed to explain the nature of line emission during the afterglow phase
of Gamma-Ray Bursts (GRBs). Trasmission and reflection models have been considered.} 
   {Given the high value ($\gg 1$) of the Thomson optical depth, the emission is likely
to arise in a reflection scenario. Within reflection models, the external reflection
geometry fails to predict the observed luminosity. On the contrary, the detected feature
can be explained in a funnel scenario with typical opening angle $\theta\sim5^{\circ}$, Nickel mass
$\sim 0.1\,\rm{M_{\odot}}$ and $T=10^{6}\,\rm{K}$. For $\theta\sim20^{\circ}$, assuming
the reprocessing material to be the SN shell, the detected emission
implies a Nickel mass $\sim0.4\,\rm{M_{\odot}}$ at $T\sim10^{7}\,\rm{K}$ and a 
metallicity $\sim10$ times the solar value.
If the giant X-ray flare
that dominates the early XRT light curve is identified with the ionizing source, 
the SN expansion began $\sim3000\,\rm{s}$ before the GRB event .
} 
   {}

   \maketitle
%

\section{Introduction}

A direct observation of Gamma-Ray Burst (GRB) central engines is not possible;
however, it is possible to infer something about the nature of their progenitors 
if indirect probes of the physical conditions, structure and composition of
the material at the GRB site can be found. X-ray spectroscopy is one of such probes.

Up to the present time, the overwhelming majority of X-ray spectra 
of afterglows detected by Swift shows a power law behaviour,
(see e.g. \citealt{OBrien})
result of a non thermal emission, the leading candidate for which is syncrotron 
emission (\citealt{Piran2005} and references therein).
Any deviation from this standard behaviour would be extremely interesting.

Detections of emission and absorption features
in addition to the basic absorbed power law  have been claimed in a
number of observations (see \citealt{Sako2005}, their Table 2
for a comprehensive summary).           
Most are interpreted as Fe K$\alpha$ emission.
The emission line radiation is supposed to be produced through
reprocessing of the burst radiation: for this reason its luminosity, 
spectral shape and wavelength shift carry important information not 
only about the geometry and structure of the surrounding medium,
but also about the physical conditions at the GRB site and the 
same GRB energetics (\citealt{Ghisellini2002}). 
However, the statistical significance of the claimed detections
has been questioned by \cite{Sako2005}: a large fraction of the claims
are based on using the $F$-test, a statistical method that
gives both false positives and false negatives when applied to emission or 
absorption features
(\citealt{Protassov2002}).
After a re-analysis of the data based on Monte Carlo simulations
\cite{Sako2005} ruled out all of the reported features.

\begin{figure*}[t!]
\centering
\includegraphics[scale=0.65]{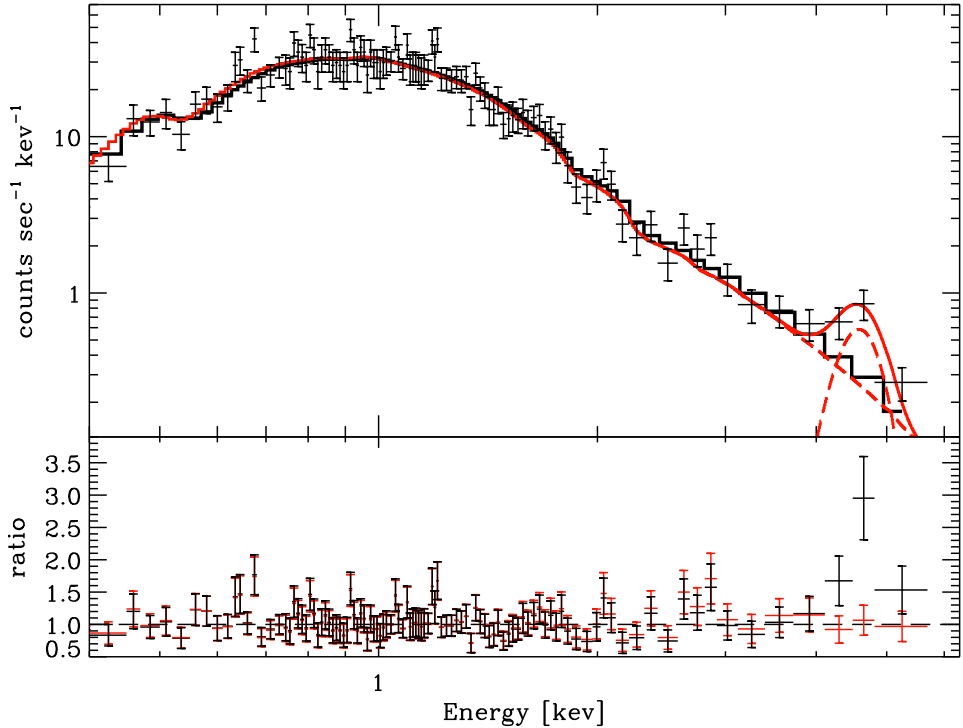}
\caption{Upper panel: GRB\,060904B spectrum observed in the
rest frame time interval $138-185\,\rm{s}$.
Solid thick black line: best fit absorbed simple power-law 
model (SPL): photon index $\Gamma=3.55^{+0.08}_{-0.08}$; 
neutral absorber column density at the source redshift 
$N_{\rm{H,z}}=(0.69^{+0.04}_{-0.04})\times10^{22}\,\rm{cm^{-2}}$.
Thin coloured line: best fit absorbed power-law plus gaussian 
model (GAU): $\Gamma=3.67^{+0.10}_{-0.09}$;
$N_{\rm{H,z}}=(0.74^{+0.05}_{-0.05})\times10^{22}\,\rm{cm^{-2}}$;
the values of the parameters related to the gaussian component
are reported in Tab. \ref{Tab:LinePar}.
The Galactic hydrogen column density $N_{\rm{H,MW}}$ has been fixed
to the value reported in \cite{Dickey}
along the GRB direction (M07).
Lower panel: ratio between the observed values and 
the model predicted values. (In color in the online edition.)}
\label{Fig:Alberto}
\end{figure*}

Recently, new detections have been claimed by \cite{Butler2007} and \cite{Moretti}
(hereafter M07).
In particular, M07 focussed on spectra of GRBs with know redshift detected by Swift in the period
April 2005-January 2007 with 
very steep spectra (photon index $\Gamma> 3$, value obtained with an absorbed single power 
law -hereafter SPL- fit). 
Out of 13 spectra with more than 2000 photons satisfying this condition,  
highly significant deviations from the SPL spectral model have been found in 4 cases. 
These spectra belong to GRB\,060502A, GRB\,060729, GRB\,060904B and GRB\,061110A.
These bursts have been detected by \emph{Swift}.
In one case (GRB\,060904B) the excess can be modeled by the emission line
of highly ionized Nickel (see Fig. \ref{Fig:Alberto}).

\begin{figure}[h!]
\centering
\includegraphics[scale=0.7 ]{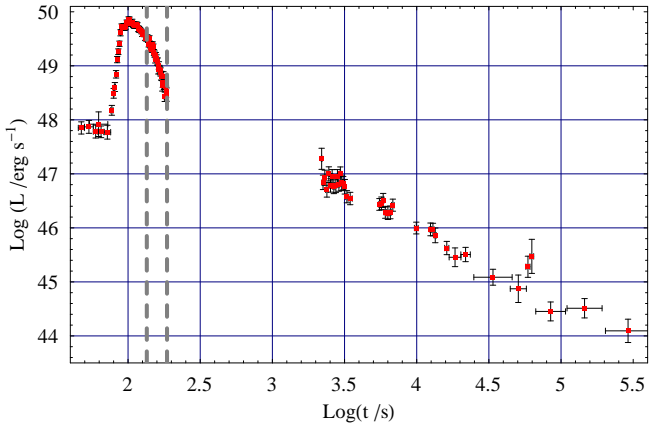}
\caption{Luminosity curve of GRB\,060904B in the 
observed $0.3-10\,\rm{keV}$ energy band (XRT observations). Rest frame times as measured from the BAT trigger
have been used. Grey dashed lines: interval of time during which the 
anomalous emission arises.}
\label{Fig:lightcurve}
\end{figure}
We refer the reader to M07 for details about the 
sample selection and the statistical methods used.

In this paper we  analyse and discuss
how and if the spectral excess detected in GRB\,060904B fits in the different models
developed to explain the production of line emission during
the afterglow phase of GRBs. 
In particular, the theoretical implications
of the possible detection of Nickel emission $\approx\,200\,\rm{s}$
(see Fig. \ref{Fig:lightcurve})
after the onset of the GRB event will be explored.    

This paper is organised as follows: the
spectroscopic identification of the line emission is discussed in
Sect. \ref{Sec:Spectroscopic}. In the same Sect.
we derive simple constraints to the theoretical models
starting from purely geometrical considerations. We discuss
the relevant physical quantities linked to the detected
emission in Sect. \ref{Sec:Physical}. Theoretical models are derived in Sect.
\ref{Sec:models}. We address the problem of the ionizing source
in Sect. \ref{ionizingsource}, while in Sect. \ref{Sec:energy} we
derive limits to the GRB total energy reservoir starting from
the detected line luminosity.
Finally, our results are discussed in Sect. \ref{Sec:discussion}.
Conclusions are drawn in Sect. \ref{Sec:Conclusion}.

The data
reduction and temporal analysis have been performed using the
standard HEADAS software, v6.1, while for the spectral
analysis we used  XSPEC (v11.3). Unless otherwise stated,
we quote errors at the 68\% of confidence level.
Through out this paper standard concordance cosmology is used:
$\Omega_{m}=0.27$, $\Omega_{\Lambda}=0.73$, $h_{0}=0.71$.  
\section{Spectroscopic identification and geometrical constraints}
\label{Sec:Spectroscopic}
\begin{table}
\begin{minipage}[t]{\columnwidth}
\caption{Best fit parameters and physical quantities related to the spectral excess
detected in GRB\,060904B X-ray afterglow.
The spectrum has been modeled with a SPL with superimposed a Gaussian  component to fit the line profile:
time interval between the GRB explosion (BAT trigger) and the detection of the spectral excess;
observed duration of the anomalous emission; Gaussian width; Gaussian central energy; 
luminosity of the additional spectral component; equivalent width. Rest frame values are provided.
(From M07).}
\label{Tab:LinePar}
\centering
\begin{center}
\begin{tabular}{l|c l}
\hline
$t_{\rm{det}}$ & 138 & (s)\\
$\Delta t$\footnote{Lower limit to the real duration of the emission because of a gap
of observation between $\sim200\,\rm{s}$ and $\sim2000\,\rm{s}$ (rest frame values)
after the GRB onset. Hence we can write $47\,\rm{s}\leq \Delta t<2000\,\rm{s}$.} & 47 & (s)\\
$\sigma_{\nu}$ &$0.50^{+0.35}_{-0.17}$&(keV)\\
$E_{0}$ &$7.85^{+0.16}_{-0.25}$ &(keV)\\
$L_{\rm{line}}$&$1.10\pm0.18$ &($10^{47}\,$ erg/s)\\
$EW$&$2.0\pm 0.3$&(keV)\\
\hline
\end{tabular}
\end{center}
\end{minipage}
\end{table} 

For GRB\,060904B, 
the statistically significant deviation from the SPL spectral model
is best modeled by an additional Gaussian component (see Fig. \ref{Fig:Alberto}).
After a Monte Carlo analysis, M07 report a multitrial 
significance of the excess greater that 99.9\%. 
We refer the reader to M07 for details.
Table \ref{Tab:LinePar} shows the physical quantities 
related to the detected excess. Unless otherwise stated, rest frame 
values will be used (the burst is located at $z=0.703$; \citealt{Fugazza}).
 
The detected excess is centered at $7.85\,\rm{keV}$ in 
agreement with what we expect 
from highly ionised Nickel line emission: the $k_{\alpha}$ 
emission of H-like (He-like) Nickel
is expected to lie at $8.10\,\rm{keV}$ ($7.81\,\rm{keV}$).
Such detection
differs from previous claims of line emission in GRB afterglows because of the
much larger luminosity involved (about 3 orders of magnitude higher; see e.g. \citealt{Bottcher}
and references therein),
and because of the really short delay between the GRB event and the line detection
($\approx10^2\,\rm{s}$). 
In previous claims of line detection the event has been reported to occur at later times, 
i.e. $10^4-10^5\,\rm{s}$ after the GRB detection.

The link between GRB events and supernova (SN) explosions
has been established in a number of observations 
(see e.g. \citealt{Galama1998}; \citealt{Stanek2003}; \citealt{Malesani2004};
\citealt{Campana}).
\emph{If} we require the line emitting material to be the SN shell,
then from the failed detection of Co features at a 
$3\sigma$ confidence level we derive an upper limit on the remnant age: $t_{\rm{remn}}\leq 6\,\rm{days}$.
(The time evolution of Ni, Co and Fe abundances as modeled by \citealt{Woosley}
has been assumed). Adopting $v_{\rm{shell}}\approx\,10^{9}\,\rm{cm\,s^{-1}}$
as typical expansion velocity of the SN shell at early stages (see e.g. \citealt{Patat}),
the maximum distance travelled by the reprocessing material is therefore
$R_{\rm{max}}\approx10^{15}\,\rm{cm}$. Assuming a contemporaneous SN-GRB
explosion this distance shrinks to $\approx 4\times 10^{12}\,\rm{cm}$. 
On the other hand, from purely geometrical considerations the detection
of the line emission at $t_{\rm{det}}$ requires the material to be
located within a distance $R= c\,t_{\rm{det}}(1-\rm{cos \phi})^{-1}$
$\approx 4\times 10^{12}(t_{\rm{det}}/138\,\rm{s})(1-\rm{cos \phi})^{-1}\,\rm{cm}$.
Note that  $R\geq 10^{16}\,\rm{cm}$ only for $\phi\leq 1^{\circ}$,
where $\phi$ is the angle between the line emitting material and
the line of sight at the GRB site.
For these reasons we consider unlikely the possibility that the detected line 
emission originates in a distant ($R\geq 10^{16}\,\rm{cm}$) reprocessor 
scenario and we prefer nearby ($R\leq 10^{13}\,\rm{cm}$) reprocessor models.
In such models the duration of the line emission is primarily linked to the
time for which the central engine is active. Unfortunately we have only
a poor lower limit to this parameter ($\Delta t \geq 47\,\rm{s}$) 
so that also geometrical factors can give important contributions to the observed 
line duration. (In the first $\approx\,130\,\rm{s}$ the high level of the
continuum emission prevents us from detecting any spectral feature with
the same luminosity even if present, while no data have been collected by
\emph{Swift} between $\sim200\,\rm{s}$ and $\sim2000\,\rm{s}$ after the onset of GRB\,060904B).
\section{Physical Conditions}
\label{Sec:Physical}
\begin{figure}[h]
\centering
\includegraphics[scale=0.7 ]{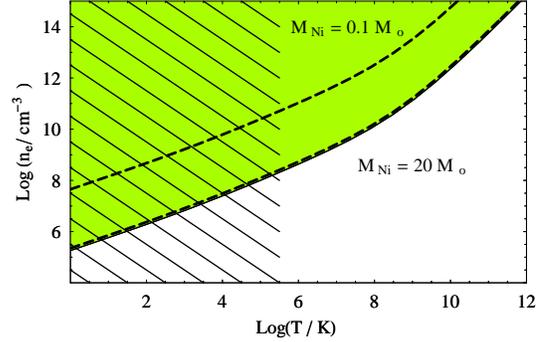}
\caption{Coloured area: electron number density-temperature parameter space  
valid for $t_{\rm{rec}}<47\,\rm{s}$ for GRB\,060904B. 
The dashed lines mark the $0.1\, \mathrm{M_{\odot}}<M_{\mathrm{Ni}}<20\,\mathrm{M_{\odot}}$
region, where $M_{\rm{Ni}}$ is the \emph{emitting} Nickel mass.
The shaded area represents the $T\leq 3.2\times 10^{5}$ K region.
The electron temperature is likely to be higher than this value 
at this stage of the SN explosion (\citealt{Arnett}).}
\label{Fig:par2}
\end{figure}

\emph{Emission Time Scale--} A line photon is produced each time
an electron recombines. The time scale of the emission process is
therefore set by the slowest between the ionization and the recombination
processes. Even under the assumption of a ionizing luminosity similar to
the luminosity of the observed excess, it can be shown that 
$t_{\rm{ion}}\ll t_{\rm{rec}}$, so that the ionization is essentially
istantaneous. The observed line luminosity implies an emission rate
of Nickel line photons of $\dot N=L_{\rm{line}}/\epsilon_{\rm{Ni}}
\approx10^{55}\,\rm{s^{-1}}$ ($\epsilon_{\rm{Ni}}\approx8\,\rm{keV}$);
if each Nickel atom recombines only once, then the emitting Nickel 
mass is given by $M_{\rm{Ni}}=56m_{\rm{p}}\dot N\Delta t$ $\approx20\,\rm{M_{\odot}}$.
\footnote{Note that in this paper we refer to $^{56}\rm{Ni}$ since this is the Ni isotope
mostly produced during SN explosions.}
However, Nickel masses as large as $\approx20\,\rm{M_{\odot}}$ are difficult to account
for in any reasonable physical scenario. We therefore conclude that each Nickel
atom recombines more than once in $\Delta t\approx47\,\rm{s}$, or $t_{\rm{rec}}<47\,\rm{s}$.
Using the expression of the recombination time of an hydrogenic ion of atomic 
number $Z=28$ given by \cite{Verner1996} the previous expression turns into a
condition on the electron density and temperature
of the plasma. The allowed parameter space is shown in Fig.
\ref{Fig:par2}.

\emph{Thomson Optical Depth--} The Thomson optical depth of the SN shell of radius
$R$ and total mass $M$ is given by:
\begin{equation}
\tau _{\mathrm{T}}\sim\frac{1}{\mu _{\mathrm{e}}m_{\mathrm{p}}}
\frac{\sigma _{\mathrm{T}}}{4\pi R^{2}}\Big(\frac{M_{\mathrm{Ni}^*}}{X}\Big)
\end{equation}
where $\mu_{\rm{e}}$ is the mean molecular weight; $X=M_{\rm{Ni}^*}/M$; $M_{\rm{Ni}^*}$ is the \emph{total}
Ni mass. The photoionization
optical depth for He-like or H-like Nickel can be written as:
\begin{equation}
\tau _{\mathrm{Ni}}=\frac{\xi\,\sigma _{\mathrm{Ni}}}{\sigma _{\mathrm{T}}}\frac{X\mu _{\mathrm{e}}}{56}\,\tau _{\mathrm{T}}
\end{equation}
where $\xi$ is the ionised to total Nickel abundance ratio and $\sigma_{\rm{Ni}}$
is the He-like or H-like Nickel photoionization cross section
($\approx10^{-20}\,\rm{cm^{-2}}$, from \citealt{Verner1995}).
It is easy to show that $\tau_{\rm{T}}\gg 1$ for $R\sim10^{12}-10^{13}\,\rm{cm}$
and $M>10^{-5}\,\rm{M_{\odot}}$, where $M$ is the total mass of the remnant shell.
While an efficient use of the H-like or the He-like Nickel ions
requires $\tau _{\mathrm{Ni}}\geq 1 $, a Thompson optical depth
larger than one is a problem: if photons diffuse through a plasma
cloud in which Thomson scattering is the predominant mechanism
and if the cloud has a central photon source, then the average
number of scatterings experienced by a photon will be $\approx1/2\,\tau _{\mathrm{T}}^{2}$
(see e.g. \citealt{Syunyaev}).
With $\tau _{\mathrm{T}}\gg 1$
, the line would be smeared in the continuum and no detection would
be possible. We must therefore require the line photons
not to have crossed such regions along our line of sight.
This in turn implies that the electron clouds 
are not isotropically distributed around the GRB location,
in order to allow the line photons to freely escape. A
$\tau _{\mathrm{T}}\gg 1$ favours models in which the line
is produced by reflection; in this case line photons come from the
thin layer of material with 
$\tau_{\rm{Ni}}\approx$ several or $\tau_{\rm{T}}\leq 1$. 
Trasmission models instead require the emitting material to \emph{globally} have
$\tau_{\rm{T}}\leq 1$.

\emph{Line Width--} In this paragraph we consider
the different physical processes that could lead to the observed
line broadening ($\approx0.5\,\rm{keV}$, Table \ref{Tab:LinePar}).
A purely thermal broadening requires $T\approx\,10^{12}\,\rm{K}$,
a temperature that is difficult to account for in any reasonable 
physical scenario. At the same time this would also cause the recombination
efficiency to drop. On the other hand, if the bulk expansion of the SN remnant is
the source of the observed broadening, then a velocity of the emitting
atoms $\approx\,25\,000\,\rm{km\,s^{-1}}$ is required. It is notable
that such velocities are indeed observed during the firts stages
of Hypernova (HN) explosions associated with GRB events (see e.g. \citealt{Patat}). 

An alternative broadening mechanism is the Compton scattering by free
electrons of the emitting material. In high temperature gas the 
first scattering will produce an average energy broadening
$\Delta E/ E\approx\,(2k_{\rm{b}}T/m_{\rm{e}}c^2)^{1/2}$ (\citealt{Syunyaev}).
For the emission feature detected in GRB\,060904B spectra, this directly 
translates into $T\approx10^7\,\rm{K}$. It is however possible that the 
real electron temperature is lower and that line photons undergo 
numerous scatterings before reaching the observer. In this case
the centroid of the line would be redshifted (line photons 
are mostly backscattered by colder electrons) with an average
energy shift given by $\Delta E\approx\,0.12\,N_{\rm{sc}}^{1/2}\,\rm{keV}$, where
the $N_{\rm{sc}}$ stands for the number of suffered scatterings (\citealt{Syunyaev}).
At the same time the multiple scattering process would further 
smear the line out, making it more difficult to detect.
It is therefore reasonable that $N_{\rm{sc}}\sim$a few.

Given the poor statistics, it is not possible to 
discuss in detail the line profile. In the following, results
will be discussed for electron temperatures in the range
$10^6-10^8\,\rm{K}$.
\section{The models}
\label{Sec:models}
In this section we discuss models in which the line is due to
reflection (see Sect. \ref{Sec:Physical}). The underlying physical mechanism
is fast ionizations and recombinations.

\begin{figure}[h]
\centering
\fbox{\includegraphics[scale=0.4]{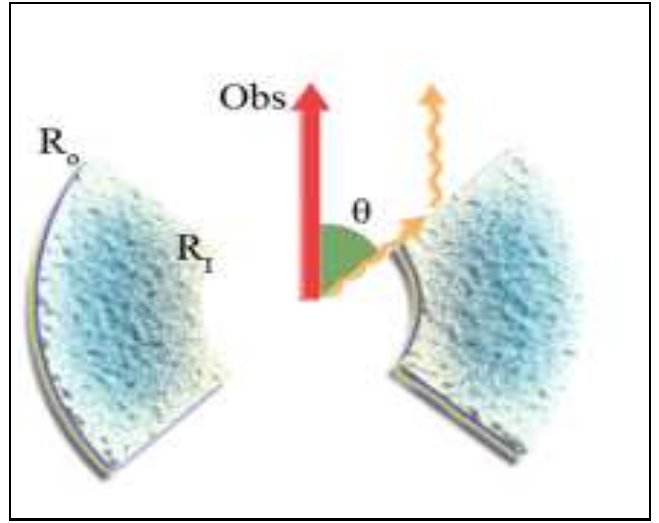}}
\caption{Sketch of the geometry assumed for the reflecting material for
the funnel model discussed in the text.}
\label{Fig:funnelmod}
\end{figure}

Reflection models (see \citealt{Bottcher} for a review) require the presence of a dense optically thick medium
surrounding the burst site. 
Moreover, an anisotropic distribution of the 
material is required in order to have a clean path where line photons
can propagate and reach the observer without being smeared in the continuum
(see Fig. \ref{Fig:funnelmod}).
The ionizing flux is efficiently reprocessed in line photons
in a superficial layer of the material with $\tau_{\rm{T}}\sim1$ 
(in order to avoid excessive Compton broadening) and $\tau_{\rm{Ni}}\sim$several,
to efficiently reprocess the continuum radiation. The volume $V_{\rm{em}}$
effectively contributing to the observed line emission is therefore given
by $V_{\rm{em}}\approx S\Delta R$ or $V_{\rm{em}}\approx S /\sigma_{\rm{T}}n_{\rm{e}}$.
The line luminosity can be expressed as:
\begin{equation}
\label{Eq:Lline}
L_{\rm{line}}=\frac{n_{\rm{Ni}}V_{\rm{em}}\epsilon_{\rm{Ni}}}{t_{\rm{rec}}}
=\frac{n_{\rm{Ni}}S\epsilon_{\rm{Ni}}\alpha_{\rm{r}}(Z,T)}{\sigma_{\rm{T}}}
\end{equation}
where:
\[
      \begin{array}{lp{0.8\linewidth}}
      \sigma_{\rm{T}}:&Thomson cross section; \\
      n_{\rm{e}}:&Electron number density;\\
      n_{\rm{Ni}}:&Ni number density;\\
      S:&Emitting surface;\\
      \epsilon_{\rm{Ni}}:& Energy of a single line photon ($\approx 8$ keV);\\
      \alpha_{\rm{r}}(Z,T):& Recombination coefficient as a function of the atomic number Z and temperature T
	 (from \citealt{Verner1996}). $[\alpha_{\rm{r}}]=\rm{cm^{3}}/\rm{s}$;\\
      \end{array}
   \]
In the following two alternative geometries of emission will be considered:
an external reflection geometry and a stratificated  funnel geometry  
(\citealt{Vietri2001}).

\subsection{External Reflection Model}
According to the external reflection model, the line emitting material
is back-illuminated by burst or afterglow photons scattered by preburst
stellar wind (we refer the reader to \citealt{Vietri2001} and references
therein for details). In this model the line emitting material has been
ejected at subrelativistic speeds along the progenitor
equator in a simultaneous SN-GRB explosion. The luminosity
$L_{\rm{scatt}}$ back-scattered by electrons moving outward from the
source is of the order of (\citealt{Vietri2001}):
\begin{equation}
L_{\rm{scatt}}\approx 1.8\times 10^{45}\frac{\dot m_{\rm{w,-5}}}{v_{\rm{w,7}}}\,\,\rm{erg\,s^{-1}}
\end{equation}
where $\dot m_{\rm{w,-5}}=\dot m_{\rm{w}}/(10^{-5}\rm{M_{\sun}yr^{-1}})$ 
and the symbol $\dot m_{\rm{w}}$ stands for the progenitor mass loss rate; 
$v_{\rm{w,7}}=v_{\rm{w}}/(10^{7}\rm{cm s^{-1}})$, being $v_{\rm{w}}$
the wind velocity.
It is notable that this is already 2 orders of magnitude \emph{lower} than
the observed line luminosity and $\approx4$ orders of magnitude lower 
than the expected ionizing luminosity: \cite{Lazzati2002} show that 
in optimal conditions a reprocessing efficiency of $\approx1\%$ at most
can be reached, so that for the emission feature detected in GRB\,060904B
$L_{\rm{ion}}\geq \,10^{49}\rm{erg\,\,s^{-1}}$ is required.
A fine-tuning of the progenitor wind velocity and of the 
mass-loss rate 
is therefore required to explain the observed line luminosity. For
this reason we consider this scenario unlikely.

\subsection{Funnel Model} In this model the line emitting material is provided by the
walls of a funnel excavated in a young SN shell (\citealt{Lazzatiphd}).
A cone geometry of opening angle $\theta$ is assumed for simplicity (see Fig. \ref{Fig:funnelmod}).
Considering a power-law electron number density profile:
\begin{equation}
n_{\rm{e}}=n_{\rm{I}}\Big ( \frac{r}{R_{\rm{I}}} \Big )^{-\alpha}
\end{equation}
equation (\ref{Eq:Lline}) translates into:
\begin{equation}
L_{\mathrm{line}}=\frac{\mu_{\mathrm{e}}X}{56\sigma_{\mathrm{T}}}
\epsilon_{\mathrm{Ni}}\,2 \pi\, \alpha_{\mathrm{r}}(Z,T)\,\rm{sin}(\theta )n_{\rm{I}}R_{\rm{I}}^{\alpha}
\int _{R_{\rm{I}}}^{R_{\rm{o}}} r^{1-\alpha}\,dr
\end{equation}
where:
\begin{equation}
n_{\rm{Ni}}=n_{\rm{e}}\frac{\mu _{\rm{e}}X}{56}=\frac{n_{\rm{e}}\mu _{\rm{e}}}{56}\frac{M_{\rm{Ni}}^{*}}{M}
\end{equation}
The total\footnote{The fraction of the total Ni mass 
actually contributing to the detected emission is indicated with $M_{\rm{Ni}}$.}
Ni mass $M_{\rm{Ni}}^{*}$ can be written as:
\begin{equation}
\label{Eq:massaNI}
M_{\rm{Ni}}^{*}=X\,M=X\,\Big( 4\pi \mu_{\rm{e}}m_{\rm{p}}\rm{cos}(\theta )n_{\rm{I}}R_{\rm{I}}^{\alpha} \int_{R_{\rm{I}}}^{R_{\rm{o}}}
r^{2-\alpha}\,dr \Big )
\end{equation}
where $m_{\rm{p}}$ is the proton mass.
This allows us to derive the line luminosity independently
of the quantity $n_{\rm{I}}R_{\rm{I}}^{\alpha}$:
\begin{equation}
\label{Eq:LLinefunnel}
L_{\rm{line}}=M_{\rm{Ni}}^{*}\frac{\epsilon_{\rm{Ni}}\,\alpha_{r}(Z,T)}{112\sigma_{\rm{T}}m_{\rm{p}}}\rm{tan}(\theta)
\frac{\int _{R_{\rm{I}}}^{R_{\rm{o}}}r^{1-\alpha} dr}{\int_{R_{\rm{I}}}^{R_{\rm{o}}}r^{2-\alpha} dr}
\end{equation}
The previous relation assumes $t_{\rm{rec}}>t_{\rm{ion}}$. A rough estimate of 
the ionization time scale $t_{\rm{ion}}$ for material located at a distance $R$
from the photon source of luminosity $L_{\rm{ion}}$ is given by:
\begin{equation}
t_{\rm{ion}}\approx\frac{\Omega R^2 \epsilon_{\rm{ion}}}{L_{\rm{ion}}\sigma_{\rm{Ni}}}
\end{equation}
where $\epsilon_{\rm{ion}}$ is the energy of the ionizing photons;
the parameter $\Omega$ allows for the possibility of collimated flux
of ionizing photons. The recombination
time scale can be written as:
\begin{equation}
t_{\rm{rec}}=\frac{1}{n_{\rm{e}}\alpha_{\rm{r}}(Z,T)}
\end{equation}
For the funnel cone geometry, the condition $t_{\rm{rec}}>t_{\rm{ion}}$
therefore becomes:
\begin{equation}
\frac{t_{\rm{rec}}}{t_{\rm{ion}}}=\frac{L_{\rm{ion}}\sigma_{\rm{Ni}}}{\Omega\epsilon_{\rm{ion}}\alpha_{\rm{r}}(Z,T)}
\frac{1}{n_{\rm{I}}R_{\rm{I}}^{\alpha}\,r^{2-\alpha}}>1
\end{equation}
As before, inserting $n_{\rm{I}}R_{\rm{I}}^{\alpha}$ from equation
(\ref{Eq:massaNI}) and assuming the collimation angle of the ionizing
photons to be $\sim\,\theta$ we finally derive:
\begin{equation}
\frac{t_{\rm{rec}}}{t_{\rm{ion}}}=\Big (\frac{L_{\rm{ion}}\sigma_{\rm{Ni}}}{\epsilon_{\rm{ion}}\alpha_{\rm{r}}}
\frac{2\rm{cos}(\theta)}{(1-\rm{cos}(\theta ))}\frac{\mu_{\mathrm{e}}m_{\rm{p}}}{M}
\frac{1}{r^{2-\alpha}}\int _{R_{\rm{I}}}^{R_{\rm{o}}} r^{2-\alpha}\, dr\Big )\,\,>1
\end{equation}
The system of equations that defines the funnel model 
is therefore:
\begin{equation}
\label{Eq:funnelsys}
   \left\{ \begin{array}{ll}
L_{\rm{line}}=M_{\rm{Ni}}^{*}\frac{\epsilon_{\rm{Ni}}\,\alpha_{r}(Z,T)}{112\sigma_{\rm{T}}m_{\rm{p}}}\rm{tan}(\theta)
\frac{\int _{R_{\rm{I}}}^{R_{\rm{o}}}r^{1-\alpha} dr}{\int_{R_{\rm{I}}}^{R_{\rm{o}}}r^{2-\alpha} dr}\\
\frac{L_{\rm{ion}}\sigma_{\rm{Ni}}}{\epsilon_{\rm{ion}}\alpha_{\rm{r}}}
\frac{2\rm{cos}(\theta)}{(1-\rm{cos}(\theta ))}\frac{\mu_{\mathrm{e}}m_{\rm{p}}}{M}
\frac{1}{r^{2-\alpha}}\int _{R_{\rm{I}}}^{R_{\rm{o}}} r^{2-\alpha}\, dr\,\,>1\\
 R_{\rm{o}}-R_{\rm{I}}\approx\,c\Delta t
\end{array}\right. 
\end{equation}
where in the last equation the duration $\Delta t$ of the line emission
is required to be comparable with the time needed by the source photons
to cross the SN shell. The observed or fixed quantities are the following:
\begin{figure}[ht!]
\centering
\includegraphics[scale=0.68]{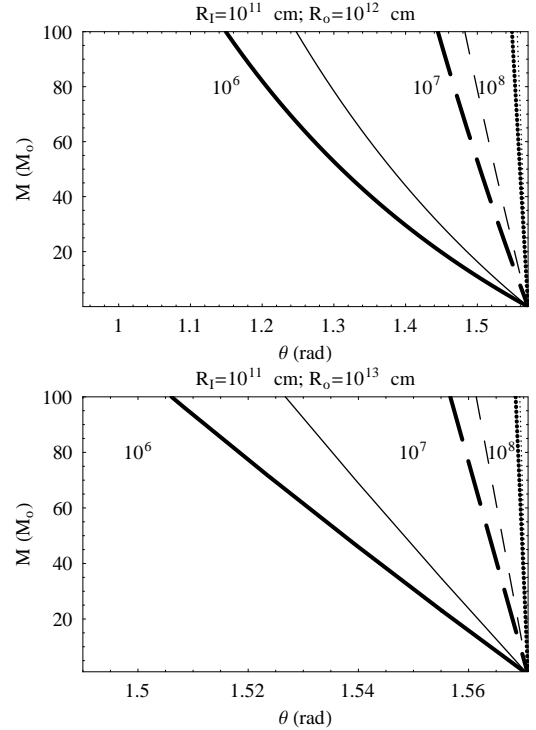}
\caption{Upper limits to the total mass of the shell derived from the
second relation of the system \ref{Eq:funnelsys} evaluated at $r=R_{\rm{o}}$
as a function of the funnel
opening angle. Solid lines: $T=10^6\,\rm{K}$; dashed lines: $T=10^7\rm{K}$;
dotted lines: $T=10^8\,\rm{K}$. Thick lines: $\alpha=0$; thin lines:
$\alpha=1$.}
\label{Figure1}
\end{figure}   
\[
      \begin{array}{lp{0.8\linewidth}}
      L_{\rm{line}}:& Line luminosity ($1.10\times10^{47}\,\rm{erg\,s^{-1}}$); \\
      \epsilon_{\rm{ion}}:& Energy of the ionizing photons (the ionization threshold is $\sim10\,\rm{keV}$); \\
    \epsilon_{\rm{Ni}}:& Energy of line photons ($\sim8\,\rm{keV}$)\\
      R_{\rm{I}}:& Internal radius of the shell, supposed to be comparable to the stellar
	progenitor radius. In order to reduce the parameter space, the system will be solved for
	$R_{\rm{I}}=10^{11}\,\rm{cm}$ (typical WR radius) and $R_{\rm{I}}=10^{12}\,\rm{cm}$ (typical blue supergiants radius).\\
	 L_{\rm{ion}}:& An efficiency of converting the illuminating continuum into the specific line
	$\sim1\%$ is assumed (\citealt{Lazzati2002}). For this reason the system will be solved for
	$L_{\rm{ion}}\sim10^{49}\,\rm{erg\,s^{-1}}$;\\
	\mu_{\rm{e}}:& Mean molecular weight. For a completely ionised material $\mu_{\rm{e}}=1,1.2$
	and $2$ for pure hydrogen, solar composition and no hydrogen respectively. In the following
	we assume $\mu_{\rm{e}}\approx2$.
      \end{array}
   \]
The unknown physical quantities of the system are:
\[
      \begin{array}{lp{0.8\linewidth}}
      \theta:& Funnel opening angle; \\
	M_{\rm{Ni}}^*:& Total Ni mass. We expect $M_{\rm{Ni}}^*\leq 1\,\rm{M_\odot}$
	(see e.g. \citealt{Arnett}).
	In particular, if the undetected SN associated with GRB\,060904B is similar to other well studied
	SNe associated with GRB events, then $M_{\rm{Ni}}^*\sim0.4\,M_\odot$ is expected (see e.g. 
	\citealt{Iwamoto}; \citealt{Mazzali2003};\citealt{Mazzali2006a};\citealt{Mazzali2006b}).
	Results will be calculated for solar metallicity $X=\,1.8\times10^{-3}$, 
	$X=\,1.8\times10^{-2}$  and 
	$X=\,1.8\times10^{-1}$;\\
	R_{\rm{o}}:& Outer radius of the shell. From the third relation of the equation \ref{Eq:funnelsys} we expect
	$R_{\rm{o}}-R_{\rm{I}}\geq 10^{12}\,\rm{cm}$;\\
	\alpha:& Mass density power-law index. The mass density of a SN shell can be approximated by
	a power-law in radius for radii greater than some critical value,
	while at radii less than critical the density is roughly constant. 
	The critical radius evolves with time; its exact value is a function of the SN energy, the 
	ejected mass and the density power law index.
	However, at early times from the SN explosion ($t<1\,\rm{hr}$) the density
	profile is nearly constant within $R_{\rm{o}}$ (see e.g. \citealt{Blondin}).
	At this distance the density undergoes
	a rapid cutoff ( see e.g. \citealt{Arnett}, their Fig. 13.2). For this reason we will consider only
	small values of the power-law index: $0\leq \alpha\leq 1$.\\
			\end{array}
   \] 
 \begin{figure*}[t]
\centering
\includegraphics[ scale=0.68 ]{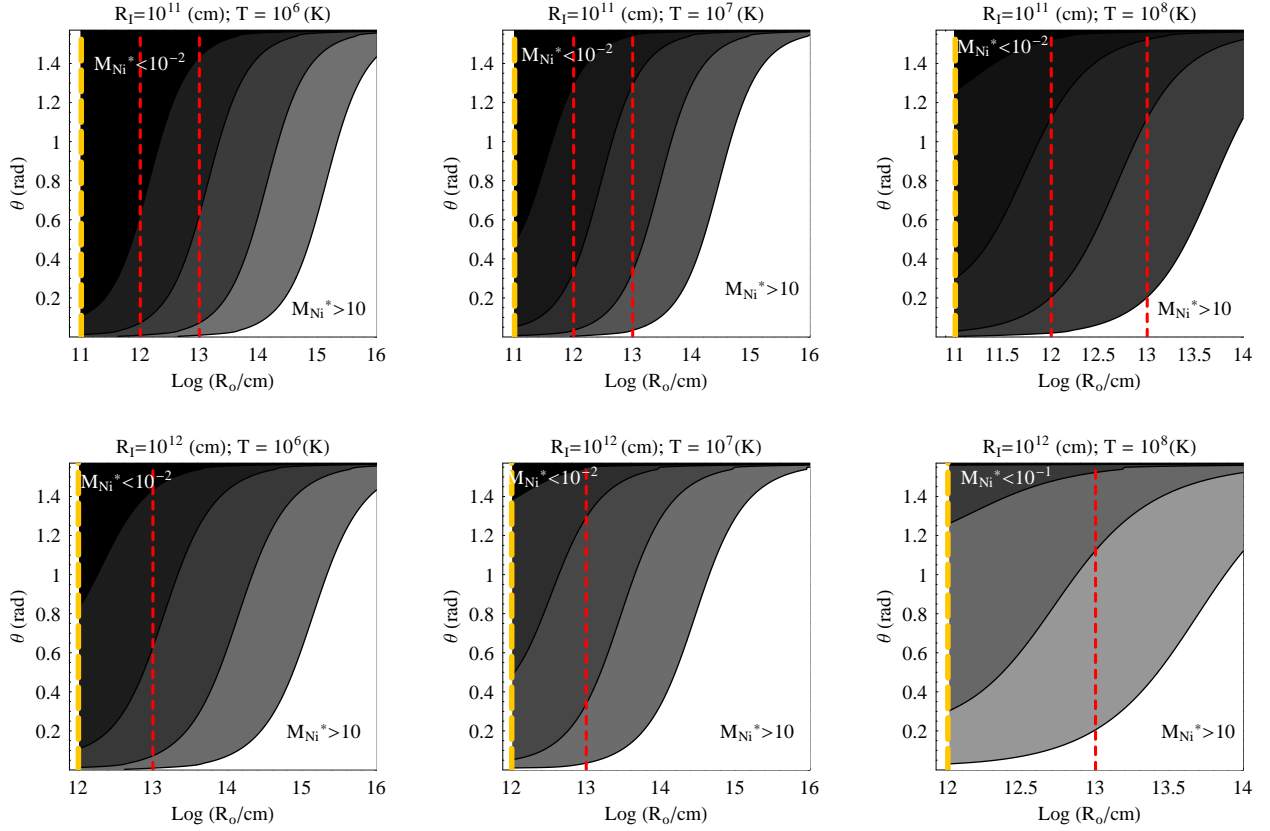}
\caption{Total Ni mass contour plot according to the first relation of the system (\ref{Eq:funnelsys}) as a function
of the funnel opening angle $\theta$ and the outer radius of the shell $R_{\rm{o}}$. 
Results are shown for the different electron temperatures indicated, $R_{\rm{I}}=10^{11}\,\rm{cm}$ and 
$10^{12}\,\rm{cm}$ and $\alpha=0$. In 
each plot, from left to right, the black solid lines mark the $M_{\rm{Ni}}^{*}=10^{-2}\rm{M_{\odot}}$,
$10^{-1}\rm{M_{\odot}}$, $1\,\rm{M_{\odot}}$ and $10\,\rm{M_{\odot}}$ region. Red dashed lines:
$R_{\rm{o}}=10^{12}$ cm and  $R_{\rm{o}}=10^{13}$ cm. $L_{\rm{line}}=1.10\times10^{47}\,\rm{erg\,s^{-1}}$,
$\epsilon_{\rm{Ni}}=8\,\rm{keV}$ have been used.}
\label{Figure2}
\end{figure*}

\begin{figure*}[t]
\centering
\includegraphics[ scale=0.75 ]{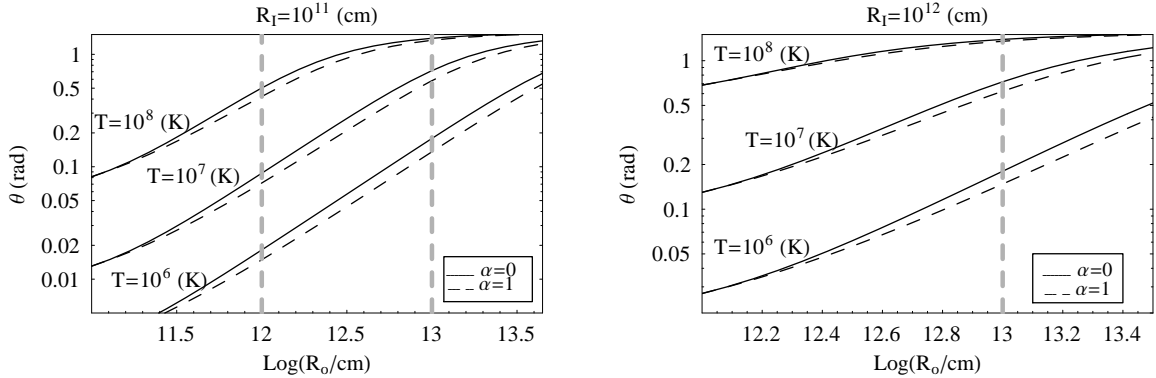}
\caption{Funnel opening angle ($\theta$) versus outer shell radius ($R_{\rm{o}}$)
according to Eq. \ref{Eq:funnelsys}, first relation. 
A total Ni mass $M_{\rm{Ni}}^*=0.4\,\rm{M_{\odot}}$ 
has been assumed (see the text for explanations). 
Results are shown
for $R_{\rm{I}}=10^{11}\,\rm{cm}$ (left panel) 
and $R_{\rm{I}}=10^{12}\,\rm{cm}$ (right panel).
In both panels the realtion is shown for three 
different temperatures: $10^6\,\rm{K}$ (bottom), $10^7\,\rm{K}$ 
and $10^8\,\rm{K}$ (top).
Solid (dashed) black lines correspond to $\alpha=0$ ($\alpha=1$). 
Dashed grey lines: $R_{\rm{o}}=10^{12},10^{13}\,\rm{cm}$.}
\label{Figure4}
\end{figure*} 

 \begin{figure*}[t]
\centering
\includegraphics[ scale=0.65 ]{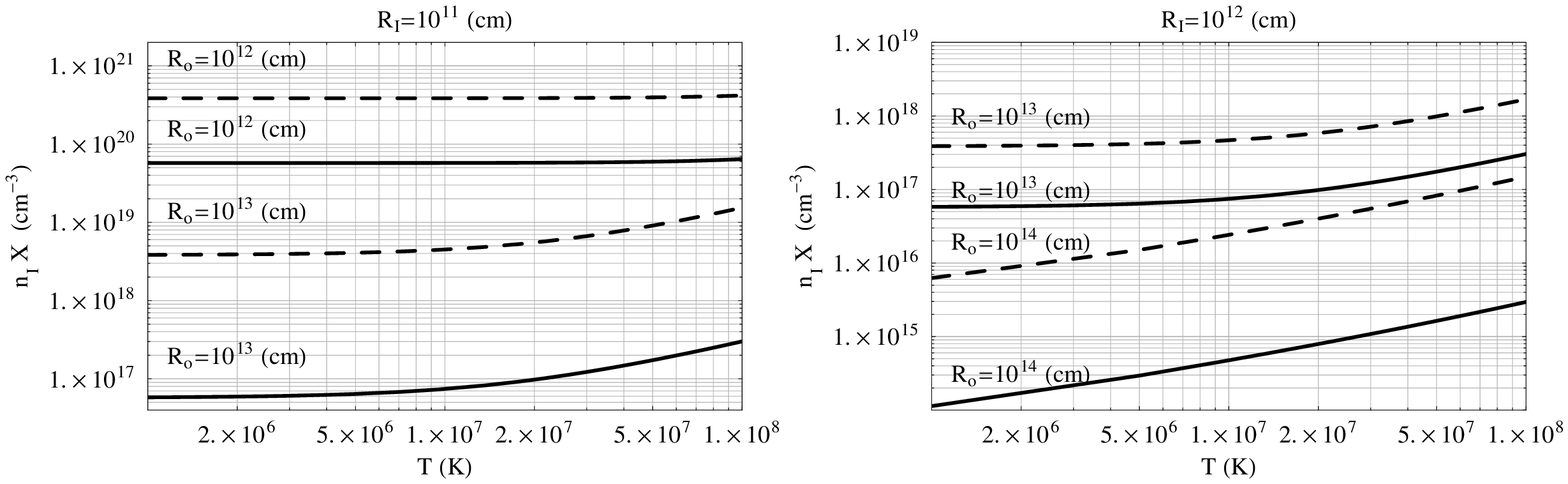}
\caption{Internal electron density $n_{\rm{I}}$ at the site of emission
times the Ni abundance ratio $X=M_{\rm{Ni}}^*/M$ as a function
of the temperature. A total Ni mass $M_{\rm{Ni}}^*=0.4\,\rm{M_{\odot}}$ has been assumed (see the text
for explanations). Left (right) panel: $R_{\rm{I}}=10^{11}\,\rm{cm}$ ($R_{\rm{I}}=10^{12}\,\rm{cm}$).
In both panels results are shown for different values of the outer radius of the SN shell $R_{\rm{o}}$. 
Solid (dashed) lines: $\alpha=0$ ($\alpha=1$).}
\label{Figure5}
\end{figure*} 

Our numerical results for the characteristic quantities of  
the problem are presented in Fig. \ref{Figure1}--\ref{Figure5}.
Figure \ref{Figure1} shows the variation of the upper limit 
to the total mass of the SN shell derived from the second
relation of the system (\ref{Eq:funnelsys}) as a function 
of the funnel opening angle. As the temperature rises,
the recombination time increases, so that at a fixed 
funnel opening angle, more Ni is needed (and 
hence more mass in the SN shell) to fulfill the 
$t_{\rm{rec}}=t_{\rm{ion}}$ requirement. Consequently
the condition $t_{\rm{rec}}>t_{\rm{ion}}$ will be satisfied
by a more extended range of SN shell masses: for example,
for $R_{\rm{I}}=10^{11}\,\rm{cm}$, $R_{\rm{o}}=10^{12}\,\rm{cm}$, $T=10^7\,\rm{K}$
we derive $M\leq 100\,\rm{M_{\odot}}$ with $\theta\leq 1.45\,\rm{rad}$.
For GRB-associated HNe, the ejected mass reported in the 
literature (see e.g. \citealt{Iwamoto}; 
\citealt{Mazzali2003};\citealt{Mazzali2006a};
\citealt{Mazzali2006b}) is however about
one order of magnitude smaller. 
For this reason we expect $M\sim 10\,\rm{M_{\odot}}$.

Figure \ref{Figure2} presents the variation of the total Ni mass
required to explain the detected emission as a function of the 
funnel opening angle and the outer radius of the shell. Results
are shown for different values of the internal radius and temperature.
The stratification index has been fixed to $0$.
As the temperature rises, the required total Ni mass increases:
this is due to the dependence of the recombination time on the
temperature. On the other hand the increase of the funnel
opening angle requires the detected emission to be explained by
a smaller amount of Ni mass. This can be understood by
considering that the density of the matter in the SN shell
is a monotonic increasing function of $\theta$. 
Consequently $t_{\rm{rec}}$ is a decreasing function of
the funnel opening angle.
The dependence of $M_{\rm{Ni}}^*$ on the matter density
manifests itself also through the parameter $R_{\rm{o}}$:
the increase of the outer shell radius requires $M_{\rm{Ni}}^*$
to increase, in order to reproduce the luminosity
of the detected emission feature. Finally, increasing 
$R_{\rm{I}}$, $M_{\rm{Ni}}^*$ decreases.

The dependence of the model on the value of the
stratification index $\alpha$ is clear from Fig. \ref{Figure4} 
and \ref{Figure5}: at a fixed outer radius, $\alpha=1$
requires smaller values of the funnel opening angle 
and higher values of the internal 
electron density. While the $\theta-R_{\rm{o}}$ relation is
nearly insensitive to the variation of this parameter,
as $\alpha$ increases from 0 to 1, the quantity $n_{\rm{I}}X$ increases
by a factor $\approx10^1-10^2$ (Fig. \ref{Figure5}).
Figure \ref{Figure4} shows the variations of the 
characteristic quantities of the problem assuming
a total Ni mass of 0.4 $\rm{M_{\odot}}$:
for $T=10^6\,\rm{K}$, $R_{\rm{I}}=10^{11}\,\rm{cm}$, 
$R_{\rm{o}}=10^{12}\,\rm{cm}$ the model requires
$\theta\approx0.02\,\rm{rad}$ ($\alpha=0$). Increasing
the electron temperature to $10^7\,\rm{K}$
($10^8\,\rm{K}$) and keeping $\alpha=0$,
the opening angle increases to 0.08 rad (0.48 rad).
The corresponding internal electron density at 
the emission site is
$n_{\rm{I}}\approx10^{20}\,\rm{cm^{-3}}$. 
The internal
electron density would be about one order of magnitude lower if
$\mu_{\rm{e}}\sim 16$, value of the mean molecular weigth 
for a SN ejecta dominated by oxygen in a low ionization stage.
\section{The ionizing source}
\label{ionizingsource}

The emission feature detected in GRB\,060904B spectra
is characterised by a large $EW=2.0\pm0.3\,\rm{keV}$.
Following \cite{Lazzati2002}, we expect an underlying 1--10 keV
continuum luminosity $L_{1-10\,\rm{keV}}\geq100\,L_{\rm{line}} $
if this is to be identified with the ionizing source.
From the \emph{Swift} XRT light curve (see Fig. \ref{Fig:lightcurve}) we derive
$L_{1-10\,\rm{keV}}=4\times10^{48}\,\rm{erg\,s^{-1}}$ at
$t\sim160\,\rm{s}$, time at which we detect the emission
feature. The \emph{detected} underlying continuum is therefore \emph{not}
physically linked to the line emission.

\begin{table}
\begin{center}
\begin{tabular}{ccccc}
\hline
 & $\Delta t$ &$\Gamma$ &$L_{\rm{1-10\,keV}}$ &$R^2 n$\\
 &            &         &                     &$\xi=10^4$\\
 & (s)& &($\rm{erg\,s^{-1}}$)&($\rm{cm^{-1}}$)\\
\hline
1 & $46-97$ & $1.61_{-0.11}^{+0.11}$&$8_{-0.5}^{+0.4}\times 10^{48}$ &$8_{-0.5}^{+0.4}\times 10^{44}$\\
2 & $97-107$ &$1.93_{-0.10}^{+0.11}$ &$4_{-0.2}^{+0.2}\times 10^{49}$ &$4_{-0.2}^{+0.2}\times 10^{45}$\\
3 & $107-117$ &$2.36_{-0.14}^{+0.15}$ &$3_{-0.2}^{+0.2}\times 10^{49}$ &$3_{-0.2}^{+0.2}\times 10^{45}$\\
4 & $117-123$ & $2.56_{-0.11}^{+0.12}$&$2_{-0.1}^{+0.1}\times 10^{49}$ &$2_{-0.1}^{+0.1}\times 10^{45}$\\
5 & $123-131$ & $2.81_{-0.12}^{+0.12}$&$2_{-0.1}^{+0.1}\times 10^{49}$ &$2_{-0.1}^{+0.1}\times 10^{45}$ \\
6 & $131-145$ &$3.15_{-0.14}^{+0.15}$ &$1_{-0.1}^{+0.1}\times 10^{49}$ &$1_{-0.1}^{+0.1}\times 10^{45}$\\
7 & $145-185$ &$3.75_{-0.17}^{+0.19}$  &$4_{-0.2}^{+0.2}\times 10^{48}$&$4_{-0.2}^{+0.2}\times 10^{44}$\\
8 & $185-64600$&$2.22_{-0.14}^{+0.15}$ &$2_{-0.3}^{+0.2}\times 10^{45}$&$2_{-0.3}^{+0.2}\times 10^{41}$\\
\hline
\end{tabular}
\caption[smallcaption]{Rest frame isotropic $1-10\,\rm{keV}$ luminosity
of GRB\,060904B as a function of time from the GRB event. The XRT data
have been splitted into different intervals of time, so that
a minimum number of $\approx2000$ photons is containd in each
spectrum. A SPL model has been assumed. $\Delta t$: interval of time;
$\Gamma$: photon index (90\% CI are listed); $R$: scale distance;
$n$:particle density; $\xi$:ionization factor (see Sect. \ref{Sec:discussion}).
For the interval of time containing the emission feature (point 7) only
the SPL contribution has been reported.}
\label{Tab:lumcon}
\end{center}
\end{table}
 
The 1--10 $\,\rm{keV}$ rest frame luminosity
as a function of time from the GRB onset is reported in Table 
\ref{Tab:lumcon}. This table shows that the detected luminosity
is actually higher than $100L_{\rm{line}}\sim10^{49}\,\rm{erg\,s^{-1}}$
in the time interval 100--130$\,\rm{s}$. Assuming that the continuum
illuminating the reprocessing material is similar to the 
observed radiation, it is therefore possible to explain the 
detected emission feature requiring the ionizing emission to be
isotropic and the reprocessing material to have a large
covering factor. In this scenario the time delay between
the arrival of direct continuum photons and the observation
of the claimed line emission $\Delta t_{\rm{D}}$ is explained in terms of light
travel time effects. 

We assume that the ionizing photons are produced at 
$t\sim100\,\rm{s}$ 
after the GRB onset or at a travelled 
distance of $R\sim3\times10^{12}\,\rm{cm}$. At this distance they
interact with the SN material.
An estimate of the funnel opening
angle $\theta$ is therefore given by:
\begin{equation}
(R+R_{\rm{p}})\,\theta\sim c \Delta t_{D}
\end{equation}
where $R_{\rm{p}}$ is the progenitor radius. With $\Delta t_{\rm{D}}\sim40\,\rm{s}$,
the previous relation translates into $\theta\sim 22^{\circ}$, $17^{\circ}$
and $5^{\circ}$ for $R_{\rm{p}}=10^{11},10^{12}$ and $10^{13}\,\rm{cm}$.  
Moreover, assuming a typical expansion velocity 
$v_{\rm{shell}}\sim10^9\,\rm{cm\,s^{-1}}$, we have:
\begin{equation}
\label{Eq:SNtime}
v_{\rm{shell}}(t+t_{\rm{SN}})\sim\,c t
\end{equation}
where $t_{\rm{SN}}$ stands for the time delay between
the GRB and the beginning of the SN expansion: for $t_{\rm{SN}}>0$
the SN event happened before the GRB explosion. For $t\sim100\,\rm{s}$
the previous relation implies $t_{\rm{SN}}\sim3000\,\rm{s}$.

\section{Limits on the total GRB energy }
\label{Sec:energy}
The emission feature detected in GRB\,060904B
shows an isotropic energy 
$E^{\rm{iso}}_{\rm{line}}\simeq L_{\rm{line}}\times\Delta t\geq 5\times10^{48}\,\rm{erg}$,
where the inequality accounts for the lower limit we have
on the duration of the emission. Following \cite{Ghisellini2002} 
the energy characterising the line emission can be used to
set lower limits on the total energy reservoir of the GRB
in the form of photons and kinetic energy:
\begin{equation}
\label{Eq:Eprima}
E=\frac{E_{\gamma}}{\eta_{\gamma}}\geq 200\frac{E_{\rm{line}}^{\rm{iso}}}{\eta_{x}\eta_{\gamma}}
\frac{\Omega_{\rm{line}}}{4\pi}
\end{equation}
where $\eta_{\rm{x}}$ is the bolometric corretion while
the symbol $\eta_{\rm{\gamma}}$ stands for the efficiency 
in converting the kinetic energy into $\gamma-$rays.
A line efficiency of $1\%$ has been assumed. For GRB\,060904B
this translates into:
\begin{equation}
\label{Eq:Eseconda}
E\geq 10^{53}\Big (\frac{0.1}{\eta_{x}}\Big )\Big (\frac{0.1}{\eta_{\gamma}}\Big )\,\,\,\rm{erg}
\end{equation}
No amplification factor $4\pi/\Omega_{\rm{line}}$ has been
applied: from Sect. \ref{ionizingsource} $\theta\geq 5^{\circ}$.
Consequently, according to \cite{Ghisellini2002} the 
amplification factor due to electron scattering is $\sim1$.
It is worth noting that the derivation of a numerical value
for the $\eta_{\rm{\gamma}}$ parameter,
depends on many
physical parameters of the burst itself and its environment:
in the conventional internal+external shock GRB model
$\eta_{\gamma}\sim1\%-5\%$ (\citealt{Zhang2004}); however, allowing
for an extremely inhomogeneous velocity of the ejecta shell
$\eta_{\gamma}\sim40\%$ (see \citealt{Kobayashi1997}; \citealt{Kobayashi2001}).
In a more recent study \cite{Zhang2007a} derive for 32 GRBs detected by 
\emph{Swift} with early X-ray afterglow data $\eta_{\gamma}<10\%$
at the end of the shallow decay phase; $\eta_{\gamma}>90\%$
at the deceleration time of the fireball. The bolometric 
correction $\eta_{x}$ is estimated using a 
Band spectrum (\citealt{Band1993}) with typical values
of the parameters: we obtain $\eta_{x}\sim0.1$.

From equations \ref{Eq:Eprima} and \ref{Eq:Eseconda}
we derive $E_{\gamma}\sim 10^{52}(0.1/\eta_{\gamma})$ erg. An independent
estimate of the isotropic equivalent radiated energy
$E_{\gamma}^{\rm{iso}}$ can be obtained as follows.
Using the correlation between the SPL photon index $\Gamma$
and the peak energy of GRB spectra measured by \emph{Swift}-BAT
in the energy band $15-150\, \rm{keV}$ (\citealt{Zhang2007b}) the photon index
of 1.70 (\citealt{GCN5520}) translates into $E_{\rm{p}}\sim85\,\rm{keV}$
or $E_{\rm{p,i}}\sim150\,\rm{keV}$ (intrinsic peak energy).
According to the Amati relation (\citealt{Amati}), the isotropic equivalent radiated energy in the 1--10000 keV
cosmological rest frame is therefore 
$E_{\gamma}^{\rm{iso}}\sim2\times10^{52}\,\rm{erg}$ 
, in good agreement with
the previous result. 

Moreover, assuming the time interval during which the ionizing source
has illuminated the reprocessing material to be comparable with
the duration of the detected emission (the recombination time
would be too small to account for the duration of the emission),
we have $E_{\rm{ill}}\sim L_{\rm{ion}}\times \Delta t\geq5\times 10^{50}\,\rm{erg} $ or
$E_{\rm{ill}}^{\rm{TOT}}\sim 2\, L_{\rm{ion}}\times \Delta t\geq 10^{51}\,\rm{erg}$.
The factor 2 in the previous relation has been added for
geometrical reasons, since we expect a two-sided geometry
with the line emitting material visible on only one side.
For GRB\,060904B this means that
a fraction of the order of $1\%$
of the radiated energy of the burst has illuminated
the line emitting material.

\section{Discussion}
\label{Sec:discussion}

The XRT light curve of GRB\,060904B is dominated by a giant flare in the
first $\approx200\,\rm{s}$ after the GRB onset (Fig. \ref{Fig:lightcurve}).
At the end of this flare the spectrum becomes softer 
and an emission feature spectroscopically
compatible with highly ionised Ni line emission arises. 

The detected emission
can be explained in a reflection scenario (Sect. \ref{Sec:models}).
In these models high electron temperatures ($\geq 10^{9}\,\rm{K}$) require
total masses of potentially 
line emitting material as high as $\sim10\,\rm{M_{\odot}}$
for $\theta\leq 60^{\circ}$. This would in turn imply a mass of the shell remnant
$\geq 100\,\rm{M_{\odot}}$ even under the assumption of 100 times solar metallicity.
For this reason, consistently with Sect. \ref{Sec:Physical} we consider unlikely
the $T\geq 10^{9}\,\rm{K}$ scenario. If the funnel opening angle $\theta$ is to be
identified with the jet opening angle $\theta_{j}$, then $\theta\sim60^{\circ}$
at $t\sim10^{2}\,\rm{s}$ after the GRB onset is difficult to justify:
\cite{Lamb2005}, starting from theoretical considerations,  
concluded that GRBs have $\theta_{j}\sim1^{\circ}$; 
from the observational point of view, a typical $\theta_{j}\sim5^{\circ}$
is derived from the afterglow jet break data (\citealt{Zhang2004b}).
If $\theta\sim5^{\circ}$ ($\sim0.1\,\rm{rad}$) the detected emission implies a total Ni
mass of the order of $0.1\,\rm{M_{\odot}}$ for 
$R_{\rm{I}}=10^{11}\,\rm{cm}$, $R_{\rm{o}}=10^{12}\,\rm{cm}$ and $M_{\rm{Ni}}^{*}\geq 1\,\rm{M_{\odot}}$
for $R_{\rm{o}}=10^{13}\,\rm{cm}$, with $M_{\rm{Ni}^{*}}$ increasing from
$T=10^{6}$ to $10^{7}\,\rm{K}$. Such high values of Ni mass
suggest that the line emitting material is a SN remnant. (Note that
up to this point no ad hoc assumption about the presence of
the SN event has been made.) The same conclusion is reached using
$R_{\rm{I}}=10^{12}\,\rm{cm}$: in this case $\theta\sim0.1\,\rm{rad}$
favours $T\sim10^{6}\,\rm{K}$ (see Fig. \ref{Figure2} and \ref{Figure4}).
We underline that in this scenario the giant flare that dominates
the early XRT ligth curve can not be identified with the ionizing 
source: in Sect. \ref{ionizingsource} we derived $\theta\geq 5^{\circ}$.
An undetected source of ionization is therefore required. 
Alternatively, non geometrical factors must be invoked
to explain the delay between the arrival of direct continuum
photons and line photons.

However, in most cases \emph{Swift} data do not support
the existence of standard jet breaks: 
using  high quality multi-wavelength data \cite{Covino2006} concluded
that no convincing case of achromatic break has been
detected in \emph{Swift} afterglows (the achromatic
behaviour was not robustly established in the pre-\emph{Swift}
era). The same conclusion has been reached by 
\cite{Panaitescu}, \cite{Willingale} and \cite{Burrows}
in independent studies.
Alternatively,
the funnel opening angle might not be directly linked to 
the jet opening angle: in other words, the naive expectation
$\theta\sim\theta_{j}$ might be wrong. 
These reasons drive us to explore the
prediction of the reflection model for larger $\theta$.
Moreover, in Sect. \ref{ionizingsource} we showed that the restarted activity
of the GRB central engine might be the source of the observed line emission
if the ionizing radiation is nearly isotropic and the material 
intercepts most of the ionizing flux. In that section we also showed that
under the assumption that the reprocessing material is the SN shell expanding 
at $v_{\rm{shell}}\sim 10^{9}\rm{cm\,s^{-1}}$, we obtain 
$\theta\sim 22^{\circ}$, $17^{\circ}$ and $5^{\circ}$ for progenitor
radii of the order of $10^{11}$, $10^{12}$ and $10^{13}\,\rm{cm}$,
respectively. In this scenario the SN shell expansion began $\sim 3000\,\rm{s}$
before the first GRB emission detected by BAT. Consequently,
at $t\sim100\,\rm{s}$ we have $R_{\rm{o}}\sim R_{\rm{p}}+3\times 10^{12}\,\,\rm{cm}$.
With $M_{\rm{Ni}}^{*}=0.4\,\rm{M_{\odot}}$, we find that the detected
emission requires $T\sim 10^{7}\,\rm{K}$ for $\theta\sim 20^{\circ}$ (0.3 rad)
regardless of the internal radius assumed. No prediction can be
made about the stratification index.

In this work we considered an efficiency of production of line photons
by the reflection mechanism of the order of $1\%$. However, 
the line production efficiency is a strong function of the
ionization parameter (\citealt{Lazzati2002}) $\xi$, here defined as:
\begin{equation}
\xi=\frac{4\pi D_{\rm{L}}^{2}F_{[1-10]}}{R^2 n} 
\end{equation}
where $F_{1-10}$ is the $1-10\,\rm{keV}$ ionizing flux; $D_{\rm{L}}$ is the
luminosity distance ($\sim4150\,\rm{Mpc}$ for $z=0.703$); $R$ is the distance 
of the reprocessing material
and $n$ is the particle density. The last column of Table \ref{Tab:lumcon},
lists the quantity $nR^2$ for $\xi=10^4$; the $1-10\,\rm{keV}$
luminosity is derived from the XRT light curve. For
a total shell mass of $10\,\rm{M_{\odot}}$ (order of magnitude of the stimated
mass ejected by HNe explosion, see e.g. \citealt{Iwamoto}; \citealt{Mazzali2003};
\citealt{Mazzali2006a};\citealt{Mazzali2006b}) and 
a scale distance $R\sim10^{12}\,\rm{cm}$ ($\sim v_{\rm{shell}}\times t_{\rm{SN}}$)
we obtain $nR^2\approx10^{45}\,\rm{cm^{-1}}$, value we actually derive
from the observed XRT luminosity between $100$ and $130\,\rm{s}$
for $\xi=10^4$. At this very high value of the ionization
parameter, Ni line production efficiencies $\approx1\%$ can be obtained
only for higher than solar metallicities (\citealt{Lazzati2002}).
In particular, the SN material associated with GRB\,060904B must have a metallicity
of the order of ten times the solar value.

Finally, GRB jets might be structured, with angle-dependent energy
per solid angle and Lorentz $\Gamma$ factor as well:
power-law jets  or Gaussian jets have been recently proposed in the literature 
(see e.g. \citealt{Zhang2002}).
Unfortunately, the jet angular structure is still unknown: for this reason
no assumption has been made about the angular distribution of the jet energy.
As a consequence, our results have been calculated assuming that the detected
radiation is representative of the whole emitted radiation, regardless of the
direction of emission. However, it is important to underline
that in the most realistic situation the emission properties of the 
ionizing continuum are likely to change with
$\theta$.

\section{Summary and Conclusions}
\label{Sec:Conclusion}

M07 recently reported the detection of a spectral feature in excess of the 
standard
single power law behaviour in 
GRB\,060904B spectra. The emission has been detected
$\approx10^2\,\rm{s}$ after the onset of the GRB, lasted more than
$47\,\rm{s}$ with an isotropic luminosity of 
$\approx\,10^{47}\,\rm{erg\,s^{-1}}$ (three orders of magnitude higher 
than the previous claimed detections). 
From the spectroscopic point of view, the detected feature
can be interpreted as He-like or H-like Ni emission. Given the poor 
statistics it is not possible to discuss the line profile.

The peculiarity of the detection motivated us to explore in detail
the theoretical implications of the presence of highly ionised Ni emitting
$\approx10^{2}\,\rm{s}$ after the burst. Given the typical $\tau_{\rm{T}}\gg 1$
(Sect. \ref{Sec:Physical}) only reflection models have been considered.
Within these models, the external reflection scenario fails to predict
the observed luminosity when standard values of the wind velocity and
mass loss rate of the progenitor star are used. On the other hand,
\emph{if} this is Ni line emission, the detected feature can be interpreted 
in a funnel scenario with the line produced in reflection and typical
Ni masses of the order of $0.1\,\rm{M_{\odot}}$ for 
$\theta=\theta_{j}\sim 5^{\circ}$, $T\sim10^{6}\,\rm{K}$.
Such high values of Ni masses strongly suggest the presence of a SN explosion
associated with the GRB event.
Alternatively, \emph{if} the reprocessing material is provided
by the SN shell and \emph{if}
the giant X-ray flare that dominates the early XRT light curve is the
ionizing source, the expansion of the SN shell began $\approx 3000\,\rm{s}$
before the GRB detection, $M_{\rm{Ni}}^*\sim0.4\,\rm{M_{\odot}}$,
$\theta\sim20^{\circ}$ and $T\sim10^{7}\,\rm{K}$.
These models imply mean electron densities $n_{\rm{e}}\sim10^{21}\,\rm{cm^{-3}}$
for ten times solar metallicity ($X\sim10^{-2}$).

In conclusion, starting from simple analytical models, we 
presented a consistent scenario in which the line emission can in
principle arise. Given the importance of the claim, the subject
certainly calls for further study.

\begin{acknowledgements}
This work is supported by OAB-INAF by ASI grant 
I/011/07/0, by the Ministry of University and Research of Italy
(PRIN 2005025417) and by the University of Milano Bicocca (Italy).
\end{acknowledgements}

\bibliographystyle{aa}
\bibliography{final}

\begin{thebibliography}{43}
\expandafter\ifx\csname natexlab\endcsname\relax\def\natexlab#1{#1}\fi

\bibitem[{{Amati}(2006)}]{Amati}
{Amati}, L. 2006, \mnras, 372, 233

\bibitem[{{Arnett}(1996)}]{Arnett}
{Arnett}, D. 1996, in {Supernova \& Nucleosynthesis}, ed. P.~U.~P.
  J.P.~Ostriker

\bibitem[{{Band} {et~al.}(1993){Band}, {Matteson}, {Ford}, {Schaefer},
  {Palmer}, {Teegarden}, {Cline}, {Briggs}, {Paciesas}, {Pendleton}, {Fishman},
  {Kouveliotou}, {Meegan}, {Wilson}, \& {Lestrade}}]{Band1993}
{Band}, D., {Matteson}, J., {Ford}, L., {et~al.} 1993, \apj, 413, 281

\bibitem[{{Blondin} \& {Ellison}(2001)}]{Blondin}
{Blondin}, J.~M. \& {Ellison}, D.~C. 2001, \apj, 560, 244

\bibitem[{{B{\"o}ttcher}(2004)}]{Bottcher}
{B{\"o}ttcher}, M. 2004, Advances in Space Research, 34, 2696

\bibitem[{{Burrows} \& {Racusin}(2007)}]{Burrows}
{Burrows}, D.~N. \& {Racusin}, J. 2007, ArXiv Astrophysics e-prints

\bibitem[{{Butler}(2007)}]{Butler2007}
{Butler}, N.~R. 2007, \apj, 656, 1001

\bibitem[{{Campana} {et~al.}(2006){Campana}, {Mangano}, {Blustin}, {Brown},
  {Burrows}, {Chincarini}, {Cummings}, {Cusumano}, {Della Valle}, {Malesani},
  {M{\'e}sz{\'a}ros}, {Nousek}, {Page}, {Sakamoto}, {Waxman}, {Zhang}, {Dai},
  {Gehrels}, {Immler}, {Marshall}, {Mason}, {Moretti}, {O'Brien}, {Osborne},
  {Page}, {Romano}, {Roming}, {Tagliaferri}, {Cominsky}, {Giommi}, {Godet},
  {Kennea}, {Krimm}, {Angelini}, {Barthelmy}, {Boyd}, {Palmer}, {Wells}, \&
  {White}}]{Campana}
{Campana}, S., {Mangano}, V., {Blustin}, A.~J., {et~al.} 2006, \nat, 442, 1008

\bibitem[{{Covino} {et~al.}(2006){Covino}, {Malesani}, {Tagliaferri},
  {Vergani}, {Chincarini}, {Kann}, {Moretti}, \& {Stella}}]{Covino2006}
{Covino}, S., {Malesani}, D., {Tagliaferri}, G., {et~al.} 2006, ArXiv
  Astrophysics e-prints

\bibitem[{{Dickey} \& {Lockman}(1990)}]{Dickey}
{Dickey}, J.~M. \& {Lockman}, F.~J. 1990, \araa, 28, 215

\bibitem[{{Fugazza} {et~al.}(2006){Fugazza}, {D'Avanzo}, {Malesani},
  {Tagliaferri}, {Chincarini}, {Stella}, {Fynbo}, {Lidman}, \&
  {Sana}}]{Fugazza}
{Fugazza}, D., {D'Avanzo}, P., {Malesani}, D., {et~al.} 2006, GRB Coordinates
  Network, 5513, 1

\bibitem[{{Galama} {et~al.}(1998){Galama}, {Vreeswijk}, {van Paradijs},
  {Kouveliotou}, {Augusteijn}, {B{\"o}hnhardt}, {Brewer}, {Doublier},
  {Gonzalez}, {Leibundgut}, {Lidman}, {Hainaut}, {Patat}, {Heise}, {in't Zand},
  {Hurley}, {Groot}, {Strom}, {Mazzali}, {Iwamoto}, {Nomoto}, {Umeda},
  {Nakamura}, {Young}, {Suzuki}, {Shigeyama}, {Koshut}, {Kippen}, {Robinson},
  {de Wildt}, {Wijers}, {Tanvir}, {Greiner}, {Pian}, {Palazzi}, {Frontera},
  {Masetti}, {Nicastro}, {Feroci}, {Costa}, {Piro}, {Peterson}, {Tinney},
  {Boyle}, {Cannon}, {Stathakis}, {Sadler}, {Begam}, \& {Ianna}}]{Galama1998}
{Galama}, T.~J., {Vreeswijk}, P.~M., {van Paradijs}, J., {et~al.} 1998, \nat,
  395, 670

\bibitem[{{Ghisellini} {et~al.}(2002){Ghisellini}, {Lazzati}, {Rossi}, \&
  {Rees}}]{Ghisellini2002}
{Ghisellini}, G., {Lazzati}, D., {Rossi}, E., \& {Rees}, M.~J. 2002, \aap, 389,
  L33

\bibitem[{{Iwamoto} {et~al.}(1998){Iwamoto}, {Mazzali}, {Nomoto}, {Umeda},
  {Nakamura}, {Patat}, {Danziger}, {Young}, {Suzuki}, {Shigeyama},
  {Augusteijn}, {Doublier}, {Gonzalez}, {Boehnhardt}, {Brewer}, {Hainaut},
  {Lidman}, {Leibundgut}, {Cappellaro}, {Turatto}, {Galama}, {Vreeswijk},
  {Kouveliotou}, {van Paradijs}, {Pian}, {Palazzi}, \& {Frontera}}]{Iwamoto}
{Iwamoto}, K., {Mazzali}, P.~A., {Nomoto}, K., {et~al.} 1998, \nat, 395, 672

\bibitem[{{Kobayashi} {et~al.}(1997){Kobayashi}, {Piran}, \&
  {Sari}}]{Kobayashi1997}
{Kobayashi}, S., {Piran}, T., \& {Sari}, R. 1997, \apj, 490, 92

\bibitem[{{Kobayashi} \& {Sari}(2001)}]{Kobayashi2001}
{Kobayashi}, S. \& {Sari}, R. 2001, \apj, 551, 934

\bibitem[{{Lamb} {et~al.}(2005){Lamb}, {Donaghy}, \& {Graziani}}]{Lamb2005}
{Lamb}, D.~Q., {Donaghy}, T.~Q., \& {Graziani}, C. 2005, \apj, 620, 355

\bibitem[{{Lazzati}(2000)}]{Lazzatiphd}
{Lazzati}, D. 2000, Phd thesis, 999

\bibitem[{{Lazzati} {et~al.}(2002){Lazzati}, {Ramirez-Ruiz}, \&
  {Rees}}]{Lazzati2002}
{Lazzati}, D., {Ramirez-Ruiz}, E., \& {Rees}, M.~J. 2002, \apjl, 572, L57

\bibitem[{{Malesani} {et~al.}(2004){Malesani}, {Tagliaferri}, {Chincarini},
  {Covino}, {Della Valle}, {Fugazza}, {Mazzali}, {Zerbi}, {D'Avanzo},
  {Kalogerakos}, {Simoncelli}, {Antonelli}, {Burderi}, {Campana}, {Cucchiara},
  {Fiore}, {Ghirlanda}, {Goldoni}, {G{\"o}tz}, {Mereghetti}, {Mirabel},
  {Romano}, {Stella}, {Minezaki}, {Yoshii}, \& {Nomoto}}]{Malesani2004}
{Malesani}, D., {Tagliaferri}, G., {Chincarini}, G., {et~al.} 2004, \apjl, 609,
  L5

\bibitem[{{Markwardt} {et~al.}(2006){Markwardt}, {Barbier}, {Barthelmy},
  {Cummings}, {Fenimore}, {Gehrels}, {Grupe}, {Hullinger}, {Krimm}, {Palmer},
  {Parsons}, {Sakamoto}, {Sato}, {Stamatikos}, \& {Tueller}}]{GCN5520}
{Markwardt}, C., {Barbier}, L., {Barthelmy}, S.~D., {et~al.} 2006, GRB
  Coordinates Network, 5520, 1

\bibitem[{{Mazzali} {et~al.}(2006{\natexlab{a}}){Mazzali}, {Deng}, {Nomoto},
  {Sauer}, {Pian}, {Tominaga}, {Tanaka}, {Maeda}, \&
  {Filippenko}}]{Mazzali2006b}
{Mazzali}, P.~A., {Deng}, J., {Nomoto}, K., {et~al.} 2006{\natexlab{a}}, \nat,
  442, 1018

\bibitem[{{Mazzali} {et~al.}(2006{\natexlab{b}}){Mazzali}, {Deng}, {Pian},
  {Malesani}, {Tominaga}, {Maeda}, {Nomoto}, {Chincarini}, {Covino}, {Della
  Valle}, {Fugazza}, {Tagliaferri}, \& {Gal-Yam}}]{Mazzali2006a}
{Mazzali}, P.~A., {Deng}, J., {Pian}, E., {et~al.} 2006{\natexlab{b}}, \apj,
  645, 1323

\bibitem[{{Mazzali} {et~al.}(2003){Mazzali}, {Deng}, {Tominaga}, {Maeda},
  {Nomoto}, {Matheson}, {Kawabata}, {Stanek}, \& {Garnavich}}]{Mazzali2003}
{Mazzali}, P.~A., {Deng}, J., {Tominaga}, N., {et~al.} 2003, \apjl, 599, L95

\bibitem[{{Moretti} {et~al.}(2007){Moretti}, {Margutti}, {Pasotti},
  {Beardmore}, {Campana}, {Chincarini}, {Covino}, {Godet}, {Guidorzi},
  {Osborne}, {Romano}, \& {Tagliaferri}}]{Moretti}
{Moretti}, A., {Margutti}, R., {Pasotti}, F., {et~al.} 2007, ArXiv e-prints,
  711

\bibitem[{{O'Brien} {et~al.}(2006){O'Brien}, {Willingale}, {Osborne}, {Goad},
  {Page}, {Vaughan}, {Rol}, {Beardmore}, {Godet}, {Hurkett}, {Wells}, {Zhang},
  {Kobayashi}, {Burrows}, {Nousek}, {Kennea}, {Falcone}, {Grupe}, {Gehrels},
  {Barthelmy}, {Cannizzo}, {Cummings}, {Hill}, {Krimm}, {Chincarini},
  {Tagliaferri}, {Campana}, {Moretti}, {Giommi}, {Perri}, {Mangano}, \&
  {LaParola}}]{OBrien}
{O'Brien}, P.~T., {Willingale}, R., {Osborne}, J., {et~al.} 2006, \apj, 647,
  1213

\bibitem[{{Panaitescu} {et~al.}(2006){Panaitescu}, {M{\'e}sz{\'a}ros},
  {Burrows}, {Nousek}, {Gehrels}, {O'Brien}, \& {Willingale}}]{Panaitescu}
{Panaitescu}, A., {M{\'e}sz{\'a}ros}, P., {Burrows}, D., {et~al.} 2006, \mnras,
  369, 2059

\bibitem[{{Patat} {et~al.}(2001){Patat}, {Cappellaro}, {Danziger}, {Mazzali},
  {Sollerman}, {Augusteijn}, {Brewer}, {Doublier}, {Gonzalez}, {Hainaut},
  {Lidman}, {Leibundgut}, {Nomoto}, {Nakamura}, {Spyromilio}, {Rizzi},
  {Turatto}, {Walsh}, {Galama}, {van Paradijs}, {Kouveliotou}, {Vreeswijk},
  {Frontera}, {Masetti}, {Palazzi}, \& {Pian}}]{Patat}
{Patat}, F., {Cappellaro}, E., {Danziger}, J., {et~al.} 2001, \apj, 555, 900

\bibitem[{{Piran}(2005)}]{Piran2005}
{Piran}, T. 2005, Reviews of Modern Physics, 76, 1143

\bibitem[{{Protassov} {et~al.}(2002){Protassov}, {van Dyk}, {Connors},
  {Kashyap}, \& {Siemiginowska}}]{Protassov2002}
{Protassov}, R., {van Dyk}, D.~A., {Connors}, A., {Kashyap}, V.~L., \&
  {Siemiginowska}, A. 2002, \apj, 571, 545

\bibitem[{{Sako} {et~al.}(2005){Sako}, {Harrison}, \& {Rutledge}}]{Sako2005}
{Sako}, M., {Harrison}, F.~A., \& {Rutledge}, R.~E. 2005, \apj, 623, 973

\bibitem[{{Stanek} {et~al.}(2003){Stanek}, {Matheson}, {Garnavich}, {Martini},
  {Berlind}, {Caldwell}, {Challis}, {Brown}, {Schild}, {Krisciunas}, {Calkins},
  {Lee}, {Hathi}, {Jansen}, {Windhorst}, {Echevarria}, {Eisenstein}, {Pindor},
  {Olszewski}, {Harding}, {Holland}, \& {Bersier}}]{Stanek2003}
{Stanek}, K.~Z., {Matheson}, T., {Garnavich}, P.~M., {et~al.} 2003, \apjl, 591,
  L17

\bibitem[{{Sunyaev} \& {Titarchuk}(1980)}]{Syunyaev}
{Sunyaev}, R.~A. \& {Titarchuk}, L.~G. 1980, A\&A, 86, 121

\bibitem[{{Verner} \& {Ferland}(1996)}]{Verner1996}
{Verner}, D.~A. \& {Ferland}, G.~J. 1996, \apjs, 103, 467

\bibitem[{{Verner} \& {Yakovlev}(1995)}]{Verner1995}
{Verner}, D.~A. \& {Yakovlev}, D.~G. 1995, \aaps, 109, 125

\bibitem[{{Vietri} {et~al.}(2001){Vietri}, {Ghisellini}, {Lazzati}, {Fiore}, \&
  {Stella}}]{Vietri2001}
{Vietri}, M., {Ghisellini}, G., {Lazzati}, D., {Fiore}, F., \& {Stella}, L.
  2001, \apjl, 550, L43

\bibitem[{{Willingale} {et~al.}(2007){Willingale}, {O'Brien}, {Osborne},
  {Godet}, {Page}, {Goad}, {Burrows}, {Zhang}, {Rol}, {Gehrels}, \&
  {Chincarini}}]{Willingale}
{Willingale}, R., {O'Brien}, P.~T., {Osborne}, J.~P., {et~al.} 2007, \apj, 662,
  1093

\bibitem[{{Woosley} \& {Weaver}(1995)}]{Woosley}
{Woosley}, S.~E. \& {Weaver}, T.~A. 1995, \apjs, 101, 181

\bibitem[{{Zhang} {et~al.}(2004){Zhang}, {Dai}, {Lloyd-Ronning}, \&
  {M{\'e}sz{\'a}ros}}]{Zhang2004b}
{Zhang}, B., {Dai}, X., {Lloyd-Ronning}, N.~M., \& {M{\'e}sz{\'a}ros}, P. 2004,
  \apjl, 601, L119

\bibitem[{{Zhang} {et~al.}(2007{\natexlab{a}}){Zhang}, {Liang}, {Page},
  {Grupe}, {Zhang}, {Barthelmy}, {Burrows}, {Campana}, {Chincarini}, {Gehrels},
  {Kobayashi}, {M{\'e}sz{\'a}ros}, {Moretti}, {Nousek}, {O'Brien}, {Osborne},
  {Roming}, {Sakamoto}, {Schady}, \& {Willingale}}]{Zhang2007a}
{Zhang}, B., {Liang}, E., {Page}, K.~L., {et~al.} 2007{\natexlab{a}}, \apj,
  655, 989

\bibitem[{{Zhang} \& {M{\'e}sz{\'a}ros}(2002)}]{Zhang2002}
{Zhang}, B. \& {M{\'e}sz{\'a}ros}, P. 2002, \apj, 571, 876

\bibitem[{{Zhang} \& {M{\'e}sz{\'a}ros}(2004)}]{Zhang2004}
{Zhang}, B. \& {M{\'e}sz{\'a}ros}, P. 2004, International Journal of Modern
  Physics A, 19, 2385

\bibitem[{{Zhang} {et~al.}(2007{\natexlab{b}}){Zhang}, {Zhang}, {Liang},
  {Gehrels}, {Burrows}, \& {M{\'e}sz{\'a}ros}}]{Zhang2007b}
{Zhang}, B., {Zhang}, B.-B., {Liang}, E.-W., {et~al.} 2007{\natexlab{b}},
  \apjl, 655, L25

\end{thebibliography}

\end{document}